\documentclass[sn-aps]{sn-jnl}

\usepackage{graphicx}%
\usepackage{multirow}%
\usepackage{amsmath,amssymb,amsfonts}%
\usepackage{amsthm}%
\usepackage{mathrsfs}%
\usepackage[title]{appendix}%
\usepackage{xcolor}%
\usepackage{textcomp}%
\usepackage{manyfoot}%
\usepackage{booktabs}%
\usepackage{algorithm}%
\usepackage{algorithmicx}%
\usepackage{algpseudocode}%
\usepackage{listings}%
\usepackage{natbib}

\UseRawInputEncoding    

\usepackage[justification=centering]{caption}   

\begin{document}

\title[Article Title]{Design, Fabrication and Characterization of Microwave Multiplexing SQUID Prototype}


\author[1,2]{\fnm{Mengjie} \sur{Song}}
\author[1,2]{\fnm{Yixian} \sur{Deng}}
\author*[1]{\fnm{Zhengwei} \sur{Li}}\email{lizw@ihep.ac.cn}
\author*[1]{\fnm{He} \sur{Gao}}\email{hegao@ihep.ac.cn}
\author[1]{\fnm{Zhouhui} \sur{Liu}}
\author[1]{\fnm{Yudong} \sur{Gu}}
\author[3]{\fnm{XiangXiang} \sur{Ren}}
\author[1]{\fnm{Nan} \sur{Li}}
\author[1]{\fnm{Guofu} \sur{Liao}}
\author[1]{\fnm{Qinglei} \sur{Xiu}}
\author[1]{\fnm{Yu} \sur{Xu}}
\author[1]{\fnm{Mengqi} \sur{Jiang}}
\author[1]{\fnm{Xufang} \sur{Li}}
\author[1]{\fnm{Yaqiong} \sur{Li}}
\author[1]{\fnm{Shibo} \sur{Shu}}
\author[1]{\fnm{Yongjie} \sur{Zhang}}
\author*[1]{\fnm{Congzhan} \sur{Liu}}\email{liucz@ihep.ac.cn}

\affil*[1]{\orgname{Key Laboratory of Particle Astrophysics, Institute of High Energy Physics, Chinese Academy of Sciences}, \orgaddress{\state{Beijing}, \postcode{100049}, \country{China}}}

\affil[2]{\orgdiv{School of Physical Science}, \orgname{University of Chinese Academy of Sciences}, \orgaddress{\city{Beijing}, \postcode{100049}, \country{China}}}

\affil[3]{\orgdiv{Key Laboratory of Particle Physics and Particle Irradiation(MOE)}, \orgname{Institute of Frontier and Interdisciplinary Science, Shandong University}, \orgaddress{\city{Qingdao, Shandong}, \postcode{266237}, \country{China}}}


\abstract{The readout system with a high multiplexing ratio has become a bottleneck limiting the application of large-scale Transition Edge Sensor (TES) detector arrays. In recent years, the microwave superconducting quantum interference device (SQUID) multiplexer has emerged as a key technology for effectively reading large-scale cryogenic detector arrays. Currently, the microwave SQUID multiplexer is being adopted by an increasing number of experiments due to its capability of achieving a multiplexing ratio of 2000:1 within the readout bandwidth. In this study, we developed a 32-channel microwave SQUID multiplexer prototype. And we measured 8 channels of the prototype. The measured equivalent noise current of the prototype reached 42 pA/$\sqrt{Hz}$.}




\maketitle

\section{Introduction}\label{sec1}

 The Transition-Edge Sensor (TES) is widely used in astronomical and cosmological observation experiments due to its extremely low noise\citep{Hubmayr-2022, Gualtieri-2016, Barret-2023}. Absorption of photon energy or power causes a temperature change in the TES, which results in a significant change of its resistance. Usually, the normal resistance of a TES detector is designed ranging from several m$\Omega$ to tens of m$\Omega$. It is hard to achieve noise impedance match with conventional semiconductor amplifiers, such as Junction Field-Effect Transistor(JFET). Superconducting Quantum Interference Device(SQUID) is nearly the only readout electronics option for TES \citep{Clarke-2004}. By reducing the noise level while increasing the number of detectors, the observation sensitivity of millimeter wave/Cosmic Microwave Background(CMB) telescopes can be rapidly enhanced. The development of TES accelerates the upgrading and replacement of millimeter wave/CMB telescopes \citep{Abazajian-2016}. The increase in the number of detectors poses a significant challenge to the low-temperature readout technology of the TES array. Therefore, a multiplexing low-temperature readout technology needs to be employed to reduce the number of readout channels. 
 
 There are four main types of multiplexing low-temperature readout technology, which are Time Division Multiplexing (TDM) \citep{Doriese-2016, Henderson-2016, Durkin-2021, Goldfinger-2024}, Frequency Division Multiplexing (FDM) \citep{Hattori-2016, Vaccaro-2024, Hartog-2018}, Code Division Multiplexing (CDM) \citep{Irwin-2010, Morgan-2016}, and microwave SQUID multiplexing($\mu$Mux) \citep{Irwin-2004, Cyndia-2023, Dober-2021,Groh-2025}. Among these techniques, TDM is the most widely employed, with a multiplexing ratio of typically around 64:1 in CMB observations\citep{Henderson-2016}. TDM utilizes DC-SQUID and incorporates at least two stages in series. However, it has stringent requirements for yield. Conversely, FDM is constrained by its limited bandwidth, with TES operating in an AC bias state. The maximum multiplexing ratio of FDM for TES X-ray microcalorimeters reaches around 70:1 \citep{Vaccaro-2024}, and it may reach 170:1 in CMB observations. CDM can achieve a significantly high multiplexing ratio theoretically. However, a primary challenge associated with CDM is its complexity. As the array size increases, the complexity of the SQUID chip design and coding also escalates \citep{Irwin-2010}. $\mu$Mux employs the principle of frequency division multiplexing and utilizes radio-frequency SQUIDs (RF-SQUIDs) instead of DC-SQUIDs, thereby eliminating the need for series connections between DC-SQUIDs. This approach results in a considerable multiplexing ratio, currently reaching 1820:1 \citep{Groh-2025}.

 $\mu$Mux has been selected as the readout electronics for various applications, including primordial gravitational wave detection and space X-ray astronomical telescopes, such as BICEP/Keck Array in South Pole \citep{Cukierman-2019}, Simons observatory in Chile \citep{Li-2020}, Ali CMB Polarization Telescope(AliCPT) in China \citep{Salatino-2020}, the conceptual X-ray space mission Lynx \citep{Bennett-2019}, etc. AliCPT is the first primordial gravitational wave experiment in China, located in Ngari(Ali), Tibet.  It has successfully achieved first light but currently operates with only one detector module. In the future, it is expected to be upgraded to 19 modules. For the upgrade of AliCPT, we designed a 32-channel $\mu$Mux chip prototype for readout development. When the fabrication technology matures, the number of readout channel for each $\mu$Mux is expected to increase to 80.

\section{Design of the microwave SQUID multiplexer}\label{sec2}
\begin{figure}[h]
    \centering
    \includegraphics[width=0.88\textwidth]{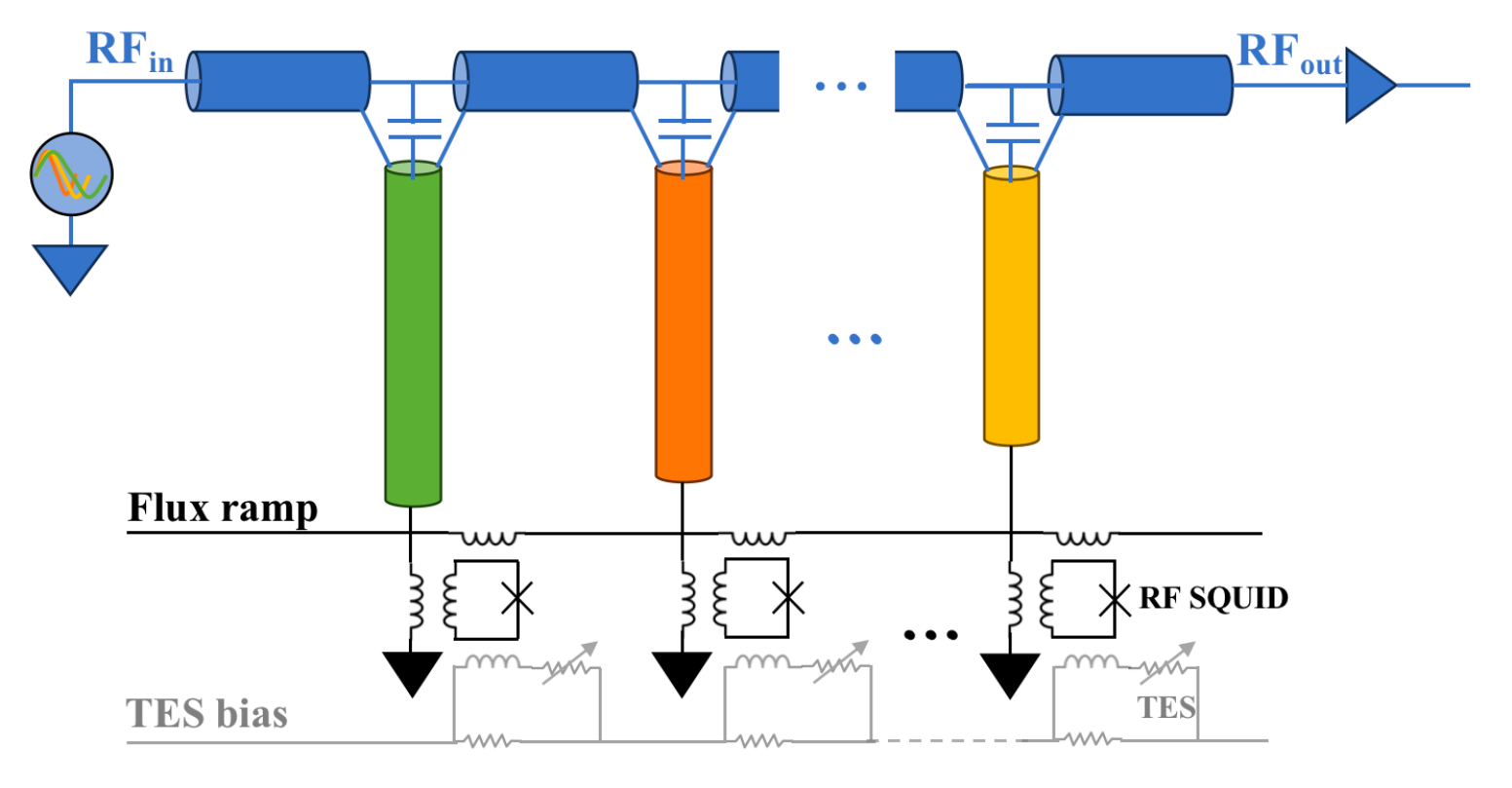}
    \caption{The equivalent electronics schematic diagram of $\mu$Mux.}
    \label{fig1}
\end{figure}

The working principle of the $\mu$Mux is illustrated in Fig.\ref{fig1}. Each input channel consists of an RF-SQUID and a coplanar waveguide (CPW) quarter-wavelength resonator. And a weak inductive coupling is adopted between the resonator and RF-SQUID, which prevents the resonator lines from crossing the SQUID wiring. The resonators connect to a common readout CPW feedline through weak capacitive coupling, using an elbow-shaped coupling method. The RF-SQUID circuit functions as a quarter-wavelength transmission line load, effectively acting as a variable inductance. The equivalent inductance of the RF-SQUID is influenced by the magnetic flux within the SQUID loop, which is modulated by the current signal flowing through the TES. Variations in the RF-SQUID's equivalent inductance result in changes in the resonant frequency and transmission parameters of the series-connected quarter-wavelength resonator. Consequently, measuring the changes in the resonator's frequency or transmission parameters can be utilized to measured the TES signal. The resonant frequencies of the quarter-wavelength resonators in each channel differ, enabling multiple channels at distinct frequencies to be read out simultaneously through the common CPW feedline. Additionally, all RF-SQUIDs are linked to a common modulation line that linearizes their responses through flux ramp modulation.

\begin{figure}[h]
    \centering
    \includegraphics[width=1\textwidth]{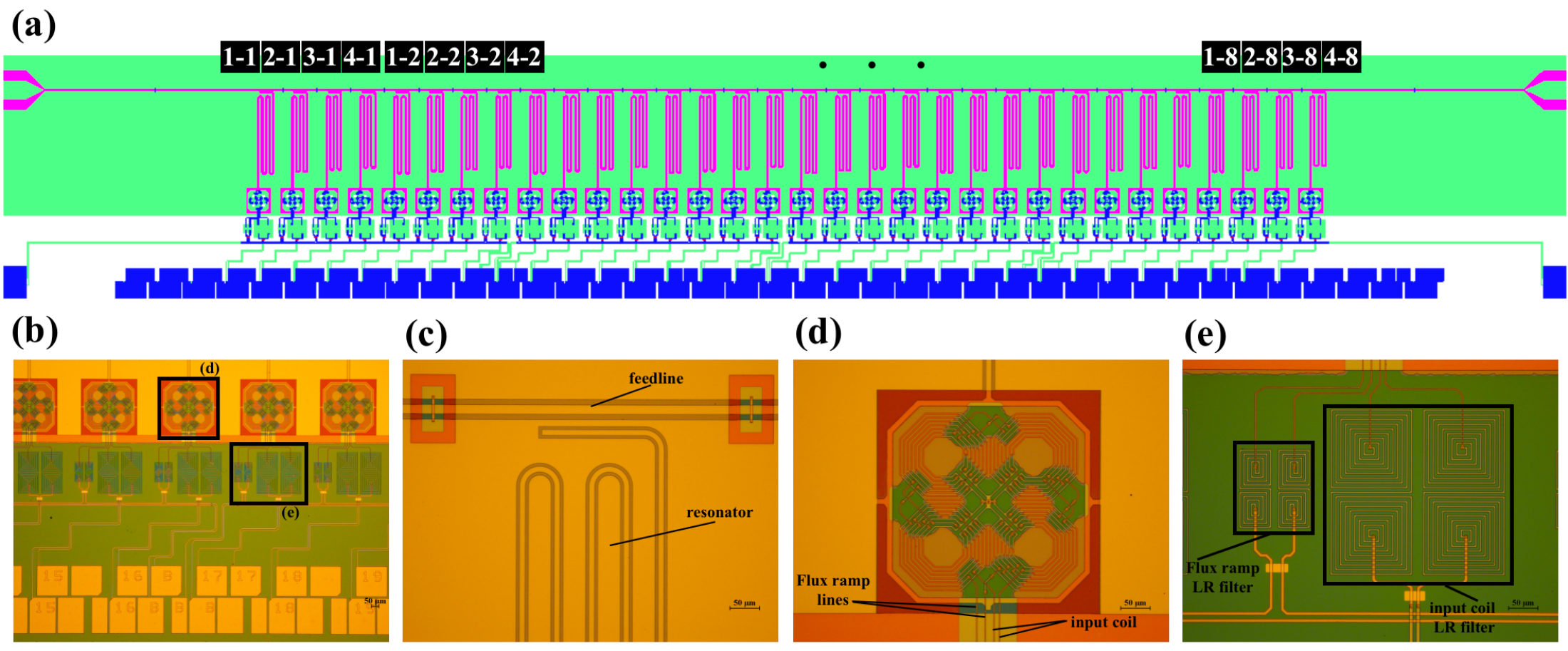}
    \caption{(a) Design of $\mu$Mux; (b) Partial microphotograph of the chip, the two black box sections are further enlarged in (d) and (e); (c) Quarterwavelength resonator with capacitively coupled to the feedline; (d) RF-SQUID and slotted washer; (e) Flux ramp LR filter (left) and input coil LR filter (right).}
    \label{fig2}
\end{figure}

Fig. \ref{fig2} presents a microphotograph of the 32-channel microwave SQUID multiplexer designed and fabricated at the Institute of High Energy Physics (IHEP), Chinese Academy of Sciences. The layout of the prototype is shown as Fig. \ref{fig2}(a). The frequency spacing for the 32 channels is set to 10 MHz. These channels are categorized into four groups, with each group comprising eight channels (e.g., 1-1, 1-2, ..., 1-8), each spaced 10 MHz apart from neighboring frequencies. The frequency interval between adjacent channels in the geometric distribution (e.g., 1-1 and 2-1, 2-1 and 3-1) is 80 MHz, which helps prevent cross-talk among different channels. The 32-channel $\mu$Mux incorporates four flux ramp lines, each capable of modulating eight RF-SQUIDs. Each flux ramp includes two corresponding bonding pads for testing. Furthermore, we have designed two bonding pads on each side of the $\mu$Mux chip. Larger-scale multiplexing is achieved by connecting these pads in series between pads. Fig. \ref{fig2}(b) illustrates the microscope image of the chip. The resonator (shown in Fig. \ref{fig2}(b)) is constructed from 200 nm thick niobium (Nb) and features a central conductor width of 9 $\mu$m. The slot width between this central conductor and the ground is 6 $\mu$m. All resonators are capacitively coupled to a common CPW feedline, which has a center conductor width of 15 $\mu$m and a gap of 10 $\mu$m from the ground. To achieve this coupling, the resonator operates parallel to the feedline CPW over a specific length (as depicted in Fig. \ref{fig2}(c)), allowing for capacitive coupling. The quality factor can be adjusted by modifying the coupling length or spacing.

\begin{table}[h]
\centering
\caption{The simulated inductances and mutual inductances of $\mu$Mux.}
\label{table1}
\begin{minipage}{0.7\textwidth}
\centering
\setlength{\tabcolsep}{0pt}
\begin{tabular*}{\linewidth}{@{\extracolsep{\fill}}lllll}
\toprule
Inductance (pH) & SQUID & Input coil & Flux ramp coil & Resonator \\
\midrule
SQUID      & 42.3   & 228.9  & 50.9   & 2.0 \\
Input coil      &        & 4144.7 & 292.3  & 18.6 \\
Flux ramp coil  &        &        & 882.6  & 5.42 \\
Resonator  &        &        &        & 161.5 \\
\botrule
\end{tabular*}
\end{minipage}
\end{table}

Each resonator is inductively coupled to the RF-SQUID. The RF-SQUID is designed as a second-order gradiometric with 4 loops to reduce magnetic interference. Each loop is an octagonal slotted washer. The linewidth of slotted-washer and coils is designed to be as narrow as 3 $\mu$m to reduce the the probability of flux-trapping. $\mu$Mux employs flux ramp modulation to linearize the input coil response.  A flux ramp coil is added to the slotted washer and the direction of the flux-ramp coil is made consistent with that of the input coil. The inductances and mutual inductances of each part are calculated based on InductEx \cite{inductex} as shown in Table \ref{table1}. The microwave excitation signal in the CPW feedline may couple inductively and enter the bias circuit of TES via the input coil.  This would cause a change in the bias state of TES and thereby affect the readout system noise. Therefore, an LR low-pass filter is added between the input coil and TES (shown as in Fig. \ref{fig2}(e)).

\section{Fabrication of the microwave SQUID multiplexer}\label{sec3}

\begin{figure}[h]
    \centering
    \includegraphics[width=0.66\textwidth]{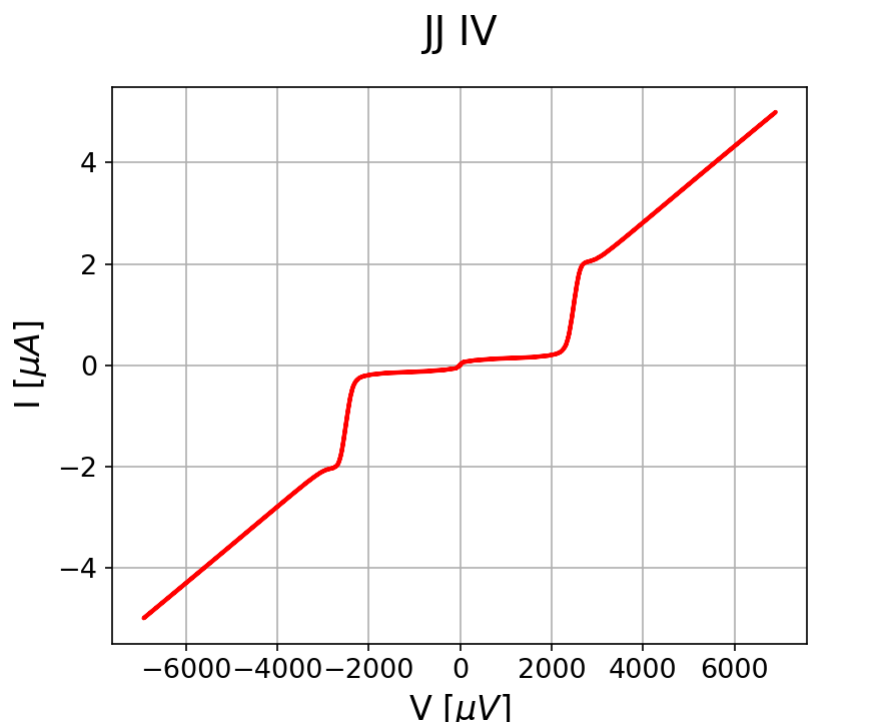}
    \caption{The measured IV curve of a 4 $\mu$m $\times$ 4 $\mu$m Josephson junction}
    \label{fig3}
\end{figure}

The $\mu$Mux prototype was fabricated at the superconducting micro-nano processing platform at IHEP. To manufacture the $\mu$Mux, we combined a high-quality Josephson tunnel junction fabrication process with the production of superconducting microwave resonators that possess high internal quality factors. Fig. \ref{fig3} presents the test results of a 4 $\mu$m $\times$ 4 $\mu$m Josephson junction created in our laboratory. The junction was measured at 3.8 K inside a pulse-tube cryostat. From the measured I-V curve, a gap current (I$_g$) of 2 $\mu$A is extracted. Using the relation I$_c$ =$\pi$ I$_g$/4, the critical current density (J$_c$) is calculated to be approximately 0.098 $\mu$A/($\mu$m)$^2$. Due to such a low J$_c$, the actual critical current (I$_c$) is incredibly small and therefore not directly discernible in the I-V curve. This suppression is primarily attributed to the high sensitivity of small critical currents to external electromagnetic interference.

\begin{figure}[h]%
    \centering
    \includegraphics[width=0.82\textwidth]{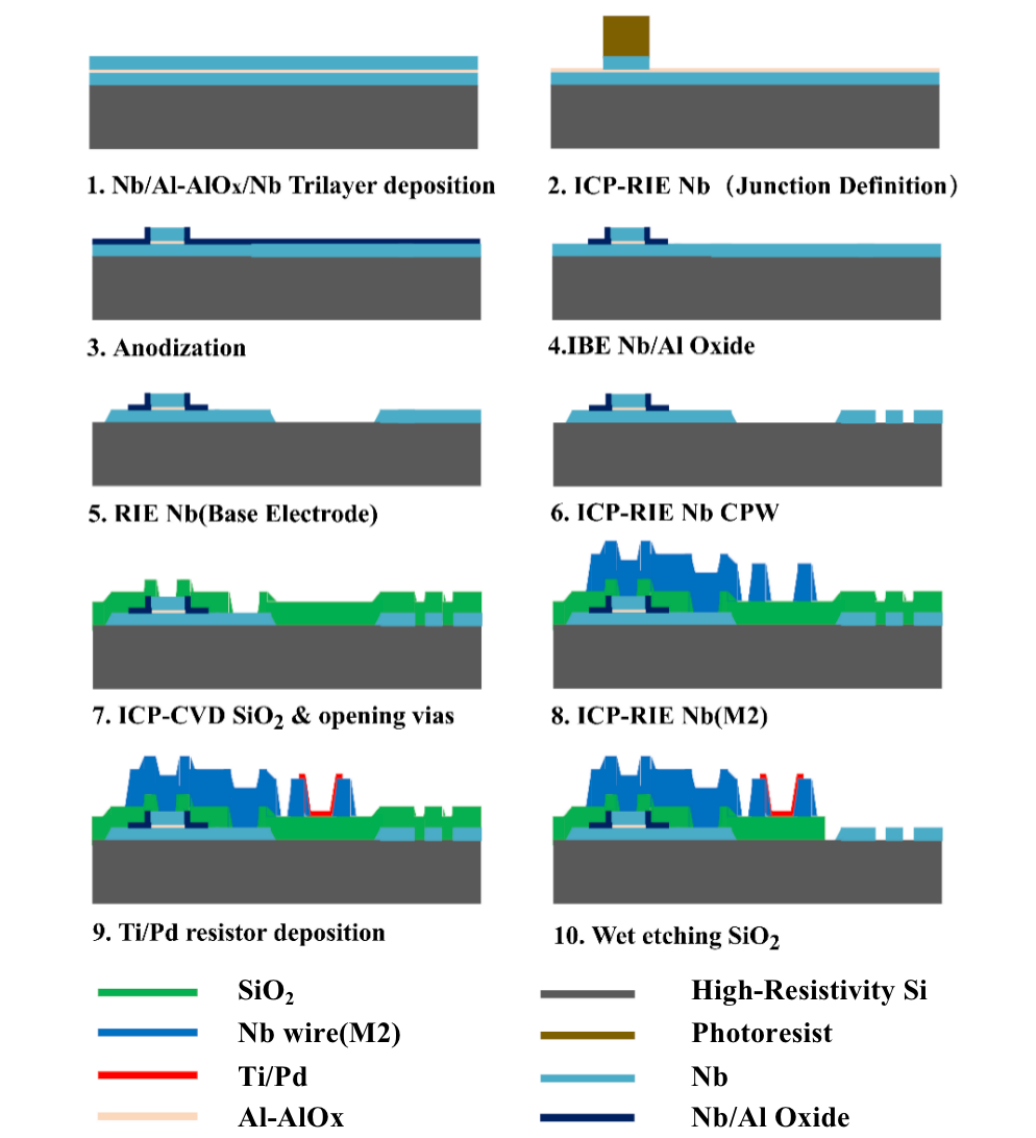}
    \caption{The fabrication procedure of $\mu$Mux}
    \label{fig4}
\end{figure}

Fig. \ref{fig4} illustrates the manufacturing process of the $\mu$Mux, which involves ten photo-lithographies, seven dry etchings, one lift-off, two wet etchings, and several deposition steps. Firstly, a trilayer film of Nb/Al-AlO$_x$/Nb (200 nm/14 nm/150 nm) was sputtered onto a 4-inch high-resistance silicon substrate using magnetron sputtering. The upper Nb layer was defined as the junction region using Inductively Coupled Plasma Reactive Ion Etching (ICP-RIE). This junction region was then protected by anodization, followed by Ion Beam Etching (IBE) of the oxide layer. The bottom superconducting Nb layer was etched in two steps. Firstly, Reactive Ion Etching (RIE) was employed to form the conductor wire. Subsequently, the Nb layer was etched again using ICP-RIE to create the CPW feedline and the quarter-wavelength CPW resonator. A 350 nm thick SiO$_2$ isolation layer was deposited using low-temperature Inductively Coupled Plasma Chemical Vapor Deposition (ICP-CVD), with a temperature of 75 degrees Celsius. Subsequently, SiO$_2$ was etched to form vias by ICP-RIE. An additional Nb layer was sputtered to act as the conductor layer, connecting the Josephson junction, input coil, flux ramp modulation coil, and LR filter, as well as leading to the pads. Next, the resistive layer was deposited by electron beam evaporation (the material was Ti/Pd). Finally, the SiO$_2$ in the gaps of the CPW and SQUID was removed through wet etching.

\section{Experimental setup for the microwave SQUID multiplexer prototype}\label{sec4}

\begin{figure}[h]%
    \centering
    \includegraphics[width=0.98\textwidth]{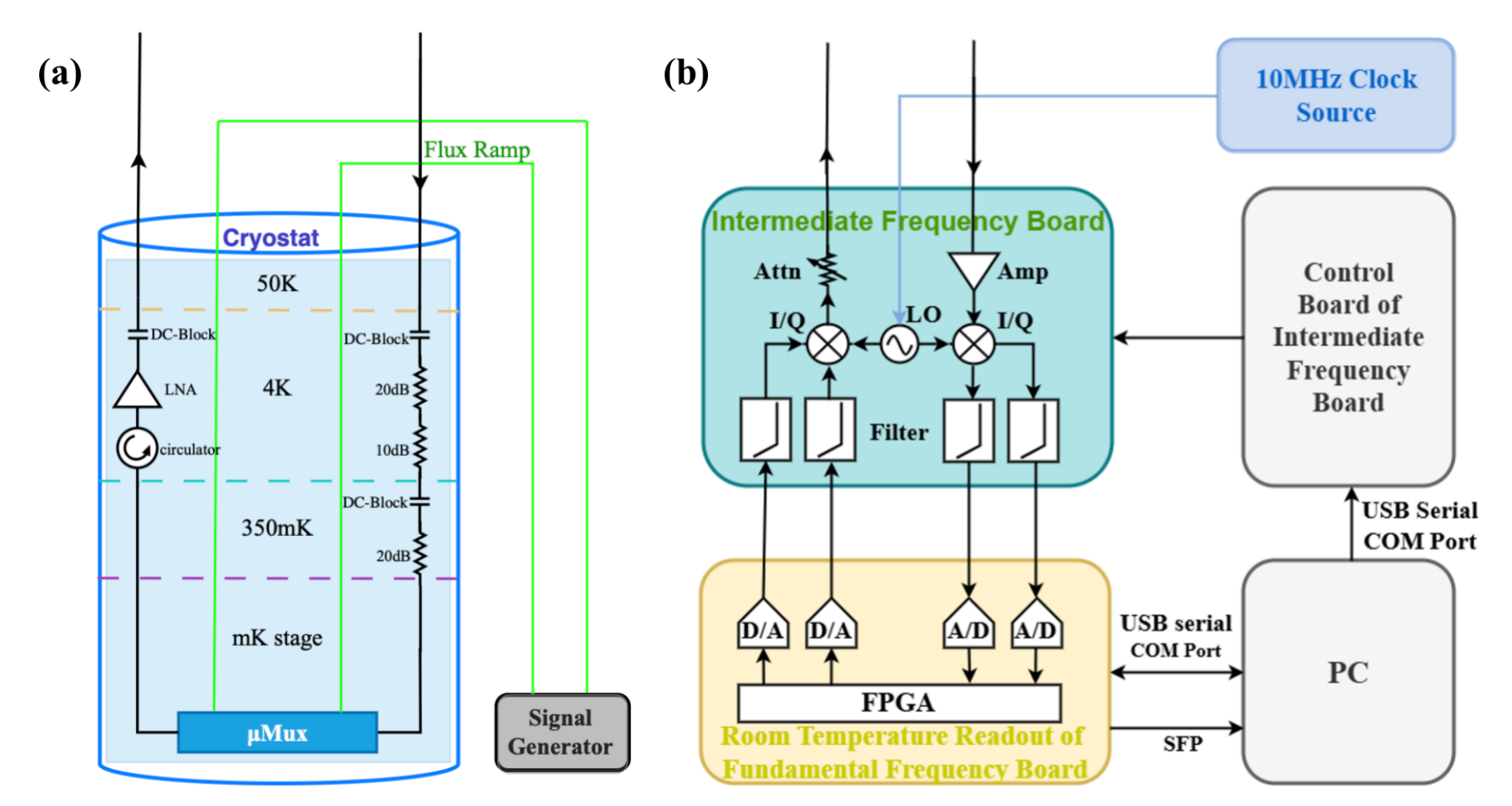}
    \caption{Block diagram of $\mu$Mux readout electronics. (a) The cryogenic readout electronics; (b) The room-temperature readout electronics.}
    \label{fig5}
\end{figure}

The overall test block diagram of the $\mu$Mux is illustrated in Fig. \ref{fig5}. This system allows us to obtain the S$_{21}$ transmission parameters and noise information of the $\mu$Mux chip. The $\mu$Mux sample is installed in an adiabatic demagnetization refrigerator (ADR), whose model is HPD 107K2. Depending on the operating temperature, the readout electronics are divided into cryogenic readout electronics located inside the refrigerator and room-temperature readout electronics situated outside. Fig. \ref{fig5}(a) depicts the cryogenic section, which includes a signal generator used to produce the flux-ramp signal. Fig. \ref{fig5}(b) illustrates the room-temperature readout electronics. The cryogenic readout electronics consist of a Low-Noise Amplifier (LNA), attenuator, circulator, and flux-ramp power supply cables. The LNA uses a High Electron Mobility Transistor (HEMT) for amplification, with a gain of approximately 32 dB. Its model is CITCRYO4-12A. The model of the low-temperature circulator is LNF-CIC4\underline{ }8A. The flux ramp power supply cables connect the flux ramp line in the chip and the signal generator.
This configuration reduces the excitation power of the comb-like spectrum generated by the room-temperature readout electronics to an appropriate level, aligning with the requirements of the $\mu$Mux resonators. Additionally, it lowers the noise temperature at the input end of the $\mu$Mux chip. The modulated comb spectrum output from the $\mu$Mux chip is then amplified by the LNA to an appropriate power level, ensuring compatibility with the readout requirements of the room-temperature electronics. The room-temperature readout section mainly comprises a fast Analog Digital Converter(ADC)/Digital Analog Converter(DAC) board, an intermediate frequency (IF) board, and an IF board control board, among other components.

The main characterization measurement of $\mu$Mux  include magnetic flux dependent measurements of the resonator's resonant frequency f$_r$, the internal quality factor Q$_i$ and the coupling quality factor Q$_c$. Therefore, we measured the transmission parameters S$_{21}$ of various magnetic fluxes passing through the SQUID loop. By changing the current of the flux ramp coil, the magnetic flux of eight channels in the same flux ramp line was simultaneously altered. We used the algorithm described in \citep{Gao-2008-thesis} to extract f$_r$, Q$_i$ and Q$_c$ from the measured data.

\section{Results}\label{sec5}

\subsection{Resonance frequency measurement}\label{subsec1}

\begin{figure}[H]%
    \centering
    \includegraphics[width=1\textwidth]{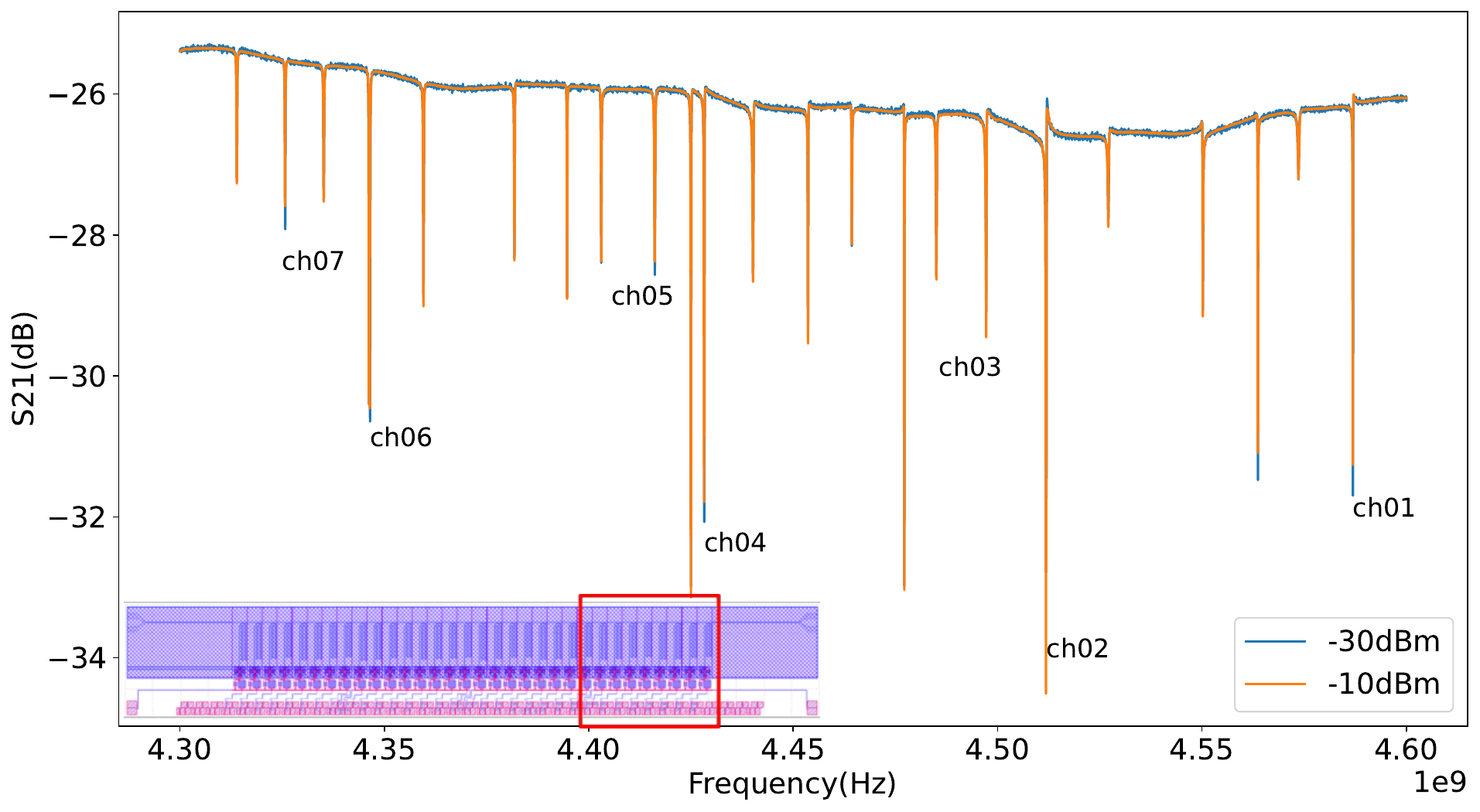}
    \caption{Measured S$_{21}$ of the $\mu$Mux prototype. The measurement was conducted at 60 mK using a vector network analyzer, with the excitation powers being -10 dBm and -30 dBm respectively.}
    \label{fig6}
\end{figure}

For the initial characterization of the microwave SQUID multiplexer, we measured the S$_{21}$ parameter of the device using a vector network analyzer for resonance frequency searching. The S$_{21}$ measurements were conducted with excitation powers of -10 dBm and -30 dBm, respectively, at a temperature of 60 mK. Fig. \ref{fig6} displays the measured transmission parameters of the microwave SQUID multiplexer prototype. The eight channels highlighted within the red box in the $\mu$Mux layout of Fig. \ref{fig6} were measured in detail. These channels are labeled as ch01 to ch07, as shown in Fig. \ref{fig6}, with the exception of one missing highest resonant frequency. When measuring the resonant frequencies of the seven channels, we performed a detailed S$_{21}$ measurement for these channels under different magnetic fluxes applied by the flux ramp coil.

\begin{figure}[h]%
    \centering
    \includegraphics[width=0.9\textwidth]{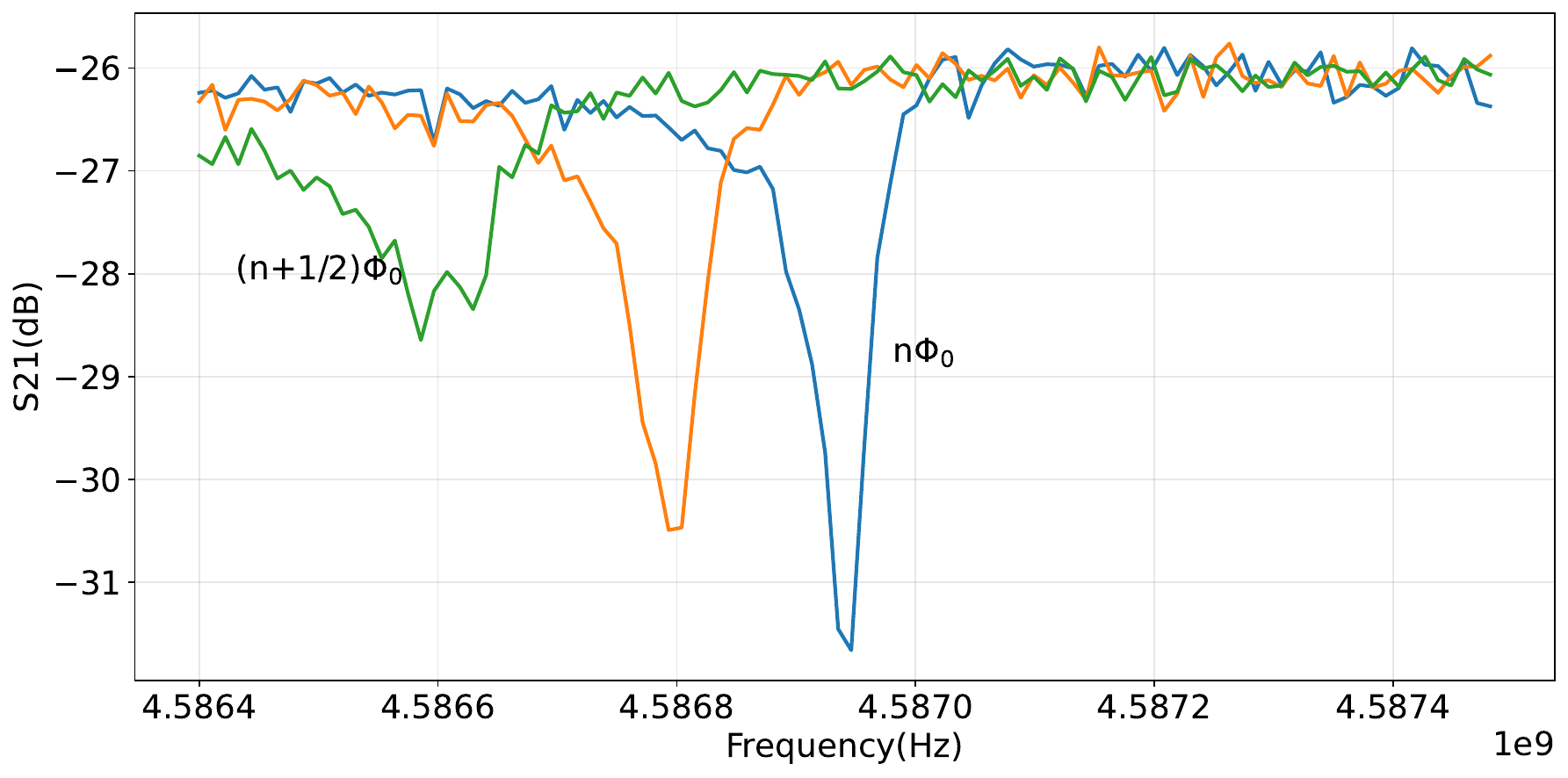}
    \caption{Measured S$_{21}$ parameters of a single channel with different magnetic fluxes in SQUID loop}
    \label{fig7}
\end{figure}

Fig. \ref{fig7} shows the S$_{21}$ parameters of a single channel under different magnetic fluxes passing through the corresponding SQUID loop, it can be seen that when the magnetic flux passing through the SQUID loop changes, the resonant frequency shifts.

\begin{figure}[h]%
    \centering
    \includegraphics[width=1\textwidth]{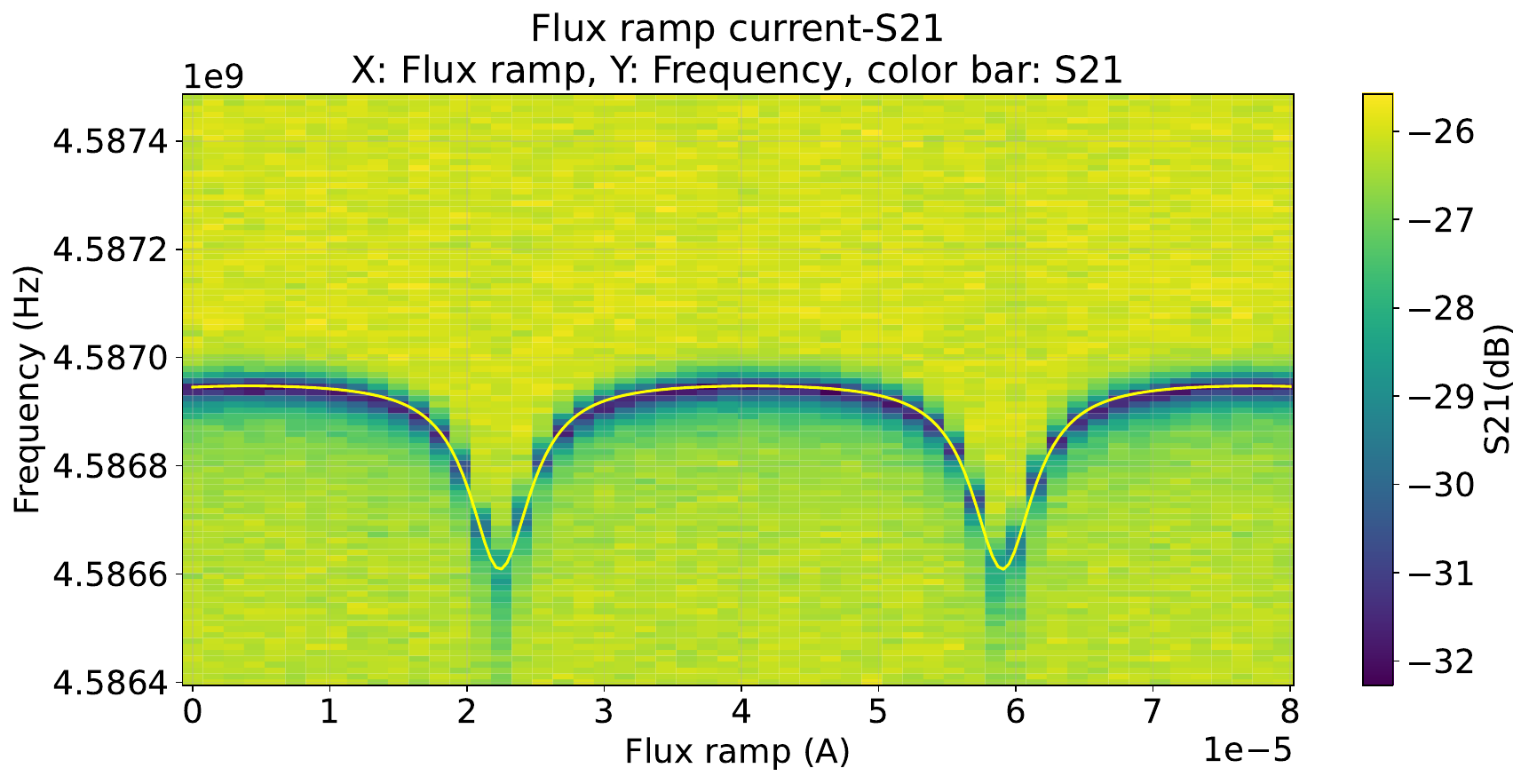}
    \caption{Measured S$_{21}$ of ch01 under different flux ramp currents. The determined resonance frequency as well as a fit(the yellow curve)  according to \citep{Gao-2008-thesis} are shown in the plot.}
    \label{fig8}
\end{figure}

\begin{figure}[H]%
    \centering
    \includegraphics[width=0.9\textwidth]{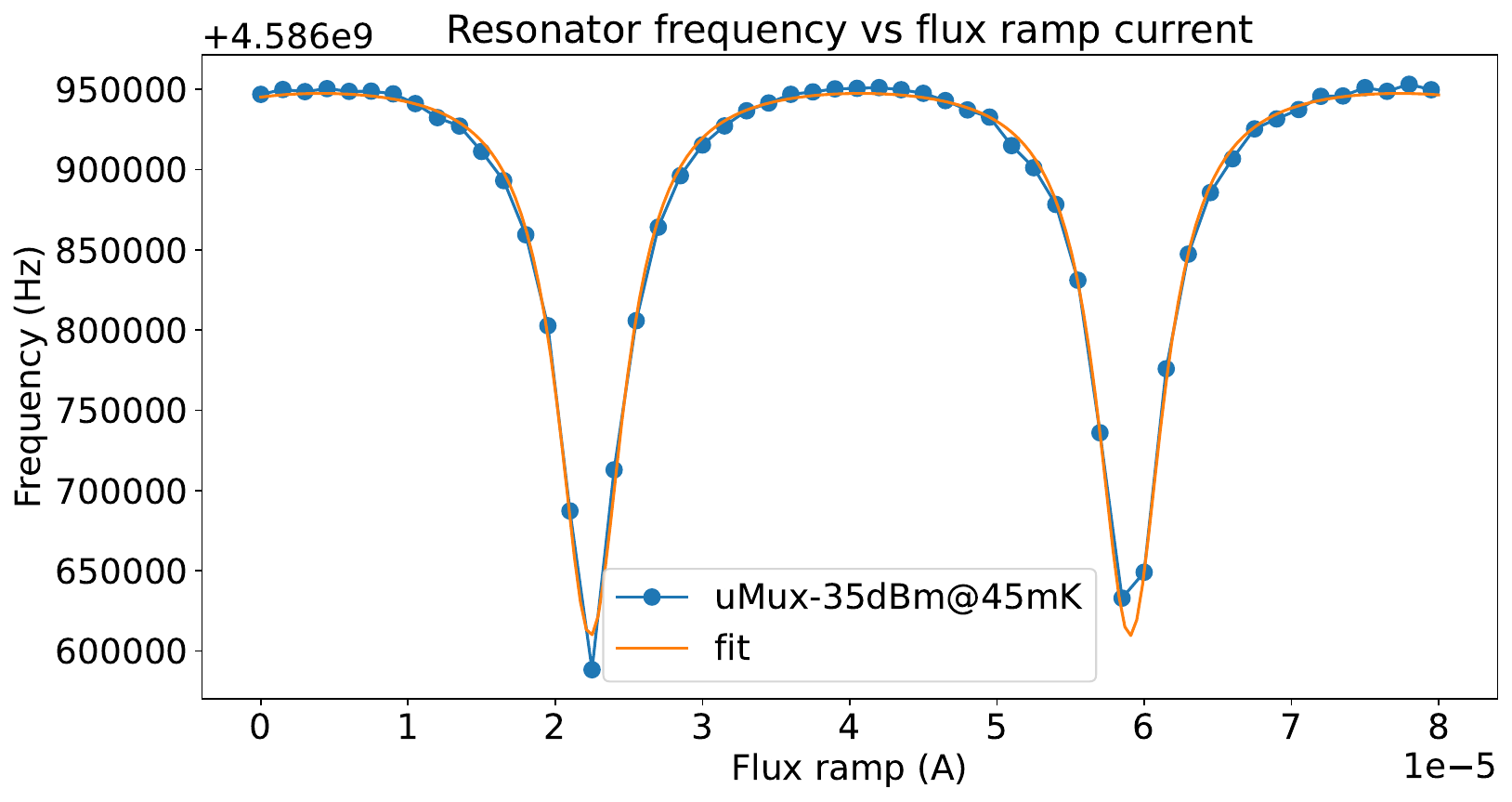}
    \caption{Dependence of resonant frequency on flux ramp current 
and curve fitting (ch01). The characterization was conducted 
at 45 mK with an excitation power of -35 dBm.}
    \label{fig9}
\end{figure}

To characterize the magnetic flux dependence of the resonant frequency f$_r$($\Phi$), different currents are applied to the flux ramp coil, generating varying magnetic fluxes in the SQUID loop. The measured S$_{21}$ parameters of ch01 under different flux ramp currents are shown in Fig. \ref{fig8}. The test was conducted at 45 mK with an excitation power of -35 dBm. Fig. \ref{fig9} illustrates the dependence of the resonant frequency on the flux ramp current (f$_r$(I$_m$$_o$$_d$)) along with the fitting of this curve based on Equation (2). Equation (1) provides the expression for the resonant frequency under the assumption that the RF-SQUID behaves as a flux-dependent inductance, f$_r$ is the unique resonance frequency of the resonator; C$_c$ is the coupling capacitor between the feedline and the resonator; Z$_0$ denotes the characteristic impedance of the resonator; L$_T$ is the load inductor that terminates the associated resonator; the mutual inductance M$_T$ represents the mutual interaction between the SQUID and the load inductor; L$_S$ is the inductance of the SQUID loop; $\lambda$ = 2$\pi$L$_S$I$_c$/$\Phi$$_0$, $\Phi$$_0$ is the flux quantum; $\phi$ = 2$\pi$$\Phi$/$\Phi$$_0$ is the normalized phase difference and depends on the magnetic flux $\Phi$ threading the SQUID loop. The relationship between the resonant frequency and flux ramp current is described by Equation (2) and derived from Equation (1), where  $\phi$ = 2$\pi$M$_{mod,eff}$I$_{mod}$/$\Phi$$_0$ + $\phi$$_{offset}$ \citep{Kempf-2017}.

\begin{equation}
f_r \approx f_0 - 4f_0^2C_CZ_0 - 
\frac{4f_0^2}{Z_0} (L_T - \frac{M_T^2}{L_S} \frac{\lambda\cos{\phi}}{1+\lambda\cos{\phi}})       
\end{equation}

\begin{equation}
f_r(\Phi) = f_{r,off} + \Delta f_{r,mod} \frac{\lambda\cos{\phi}}{1+\lambda\cos{\phi}}  
\end{equation}

Here, $\phi_{\text{offset}}$ represents the normalized magnetic flux offset passing through the SQUID loop, $M_{\text{mod}, \text{eff}}$ refers to the effective mutual inductance between the SQUID and the modulated coil. Due to parasitic coupling between the flux ramp coil and the input coil, we utilize the effective mutual inductance $M_{\text{mod}, \text{eff}}$. The fitting results accurately describe our measurement outcomes. We can determine the flux-independent term in the resonant frequency expression $f_{r, \text{off}} \equiv f_0 - 4f_0^2 C_C Z_0 - \frac{4f_0^2 L_T}{Z_0}$, the modulation coefficient $\Delta f_{r, \text{mod}} \equiv \frac{4f_0^2 M_T^2}{Z_0 L_S}$, the hysteresis parameter $\lambda$, and the effective mutual inductance $M_{\text{mod}, \text{eff}}$. We conducted this analysis for all seven measured channels. Table \ref{table2} summarizes the relevant parameters obtained from the fitting of these channels. The design value of the resonance frequency is within the range of 4.251-4.509 GHz. The results show that both the measured resonant frequency and the corresponding design value deviate by approximately 80 MHz, which may be attributed to manufacturing errors. The measured values of the effective mutual inductance for different channels are generally consistent, with a deviation from simulation value(50.9 pH) within 12\%. The design value of hysteresis parameter $\lambda$ is approximately 0.6, but the result ranges from 0.6 to 0.9, appearing somewhat random. Using the calculated value of the shielded SQUID inductance $L_S$, we find the critical current to range from 4.6 to 6.9 $\mu$A. This is close to our expected critical current, though it shows variability. This variability may arise from unideal processing uniformity. 

\begin{table}[h]
\centering
\caption{Summary of fitting parameters for the measured 7 channels.}
\label{table2}
\begin{minipage}{0.7\textwidth}
\centering
\setlength{\tabcolsep}{0pt} 
\begin{tabular*}{\linewidth}{@{\extracolsep{\fill}}llll}
\toprule
Channel & $f_r$/GHz & $\lambda$ & $M_{\rm mod,eff}$/pH \\
\midrule
1 & 4.5869 & 0.901 & 56.42 \\
2 & 4.5120 & 0.618 & 56.79 \\
3 & 4.4973 & 0.913 & 56.35 \\
4 & 4.4283 & 0.909 & 56.39 \\
5 & 4.4162 & 0.913 & 56.21 \\
6 & 4.3465 & 0.707 & 56.26 \\
7 & 4.3352 & 0.714 & 56.36 \\
\botrule
\end{tabular*}
\end{minipage}
\end{table}

\subsection{Quality factor measurement}\label{subsec2}

\begin{figure}[h]%
    \centering
    \includegraphics[width=0.9\textwidth]{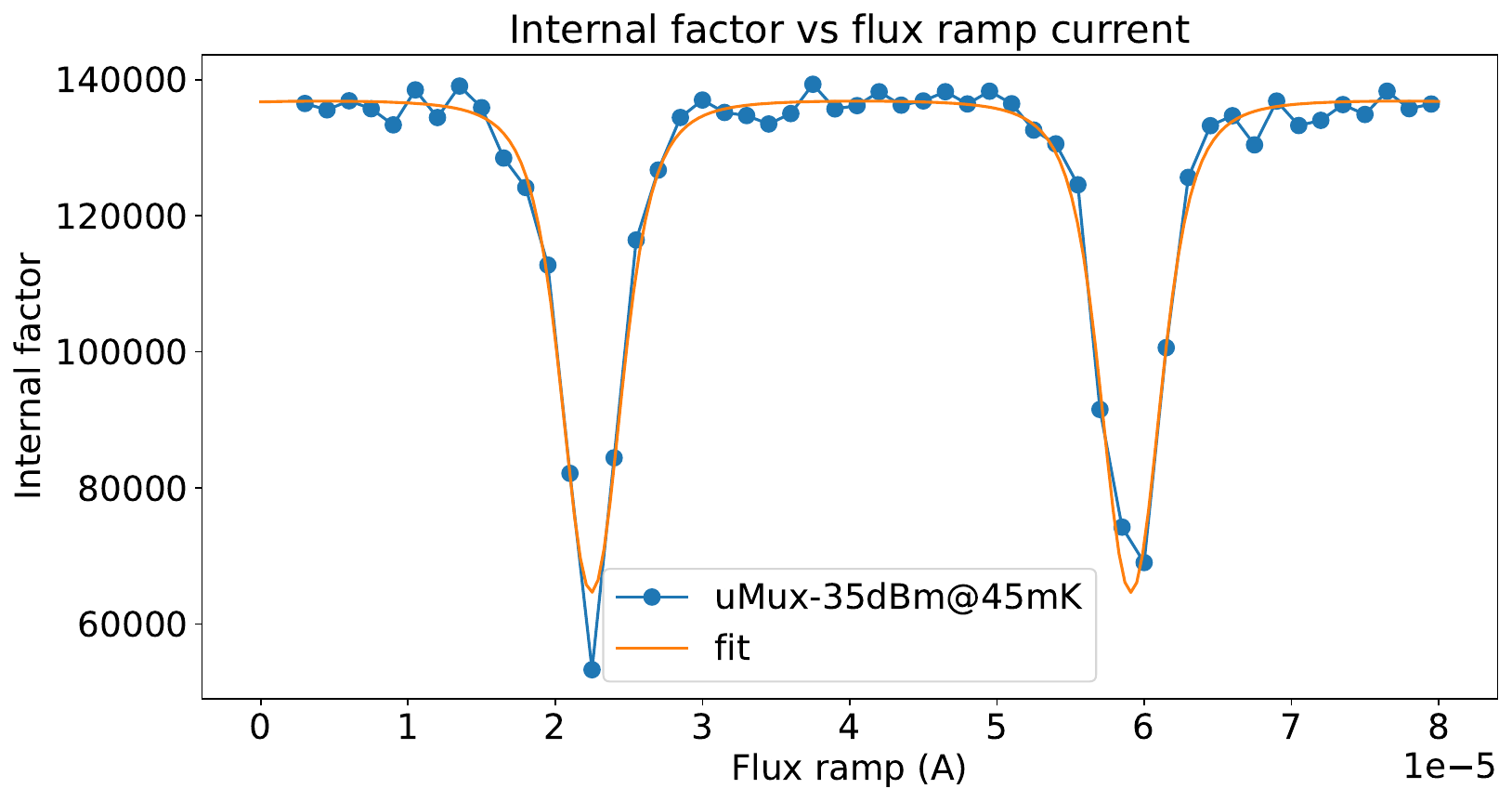}
    \caption[width=0.9\textwidth]{The dependence of the internal quality factor Q$_i$ of ch01 on the flux ramp current. The measurement was carried out at a temperature of 45 mK, using an excitation power of -35 dBm. The solid orange line is the fitting based on formula (3), assuming that the internal quality factor is affected by the sub-gap resistance of the Josephson junction.}
    \label{fig10}
\end{figure}

The internal quality factor of the microwave resonator is related to the magnetic flux $\Phi$ passing through the SQUID loop. To investigate this effect, Fig. \ref{fig10} presents the relationship between the measured internal quality factor $Q_i$, extracted from ch01 during the resonance curve fitting process, and the flux ramp current. The tests were conducted at 45 mK with an excitation power of -35 dBm. It is evident that $Q_i$ and the flux ramp current exhibit a periodic dependence, with the period matching that of the resonant frequency's dependence on the flux ramp current. According to the multiplexer model, periodic modulation of the flux ramp current is expected when the sub-gap resistance of the Josephson junction is small. Therefore, we use the following formula for fitting:

\begin{equation}
\frac{1}{Q_i} = \frac{1}{Q_{\text{offset}}} + \left(\frac{\pi Z_0 R_{sg} (1+\lambda \cos\phi)^2}{4(\omega M_T)^2}\right)^{-1}
\end{equation}

Here,  Z$_0$, $\lambda$, $\phi$ and M$_T$ are the same as in equation (1). $Q_{\text{offset}}$ represents the offset of the internal quality factor, incorporating all contributions that are independent of the magnetic flux. $R_{sg}$ is the sub-gap resistance of the Josephson junction\citep{Kempf-2017}. During the fitting process, $Q_{\text{offset}}$ and $R_{sg}$ are the only fitting parameters, as we fixed $\lambda$ to the value obtained from fitting the dependence of the resonant frequency on the flux ramp current. It can be observed that there is a strong consistency between the measured internal quality factor and the fitted model. Based on the test results of our seven channels, the value of the sub-gap resistance $R_{sg}$ ranges from 60 $\Omega$ to 1160 $\Omega$. This is significantly lower compared to the sub-gap resistance value of high-quality Nb/Al-AlO$_x$/Nb junctions. This discrepancy is likely due to deviations in the optimal sputtering parameters necessary for achieving low intrinsic stress in the Josephson junction or damage caused by the etching process. We remain optimistic about resolving this issue in the future.

\begin{figure}[H]%
    \centering
    \includegraphics[width=0.9\textwidth]{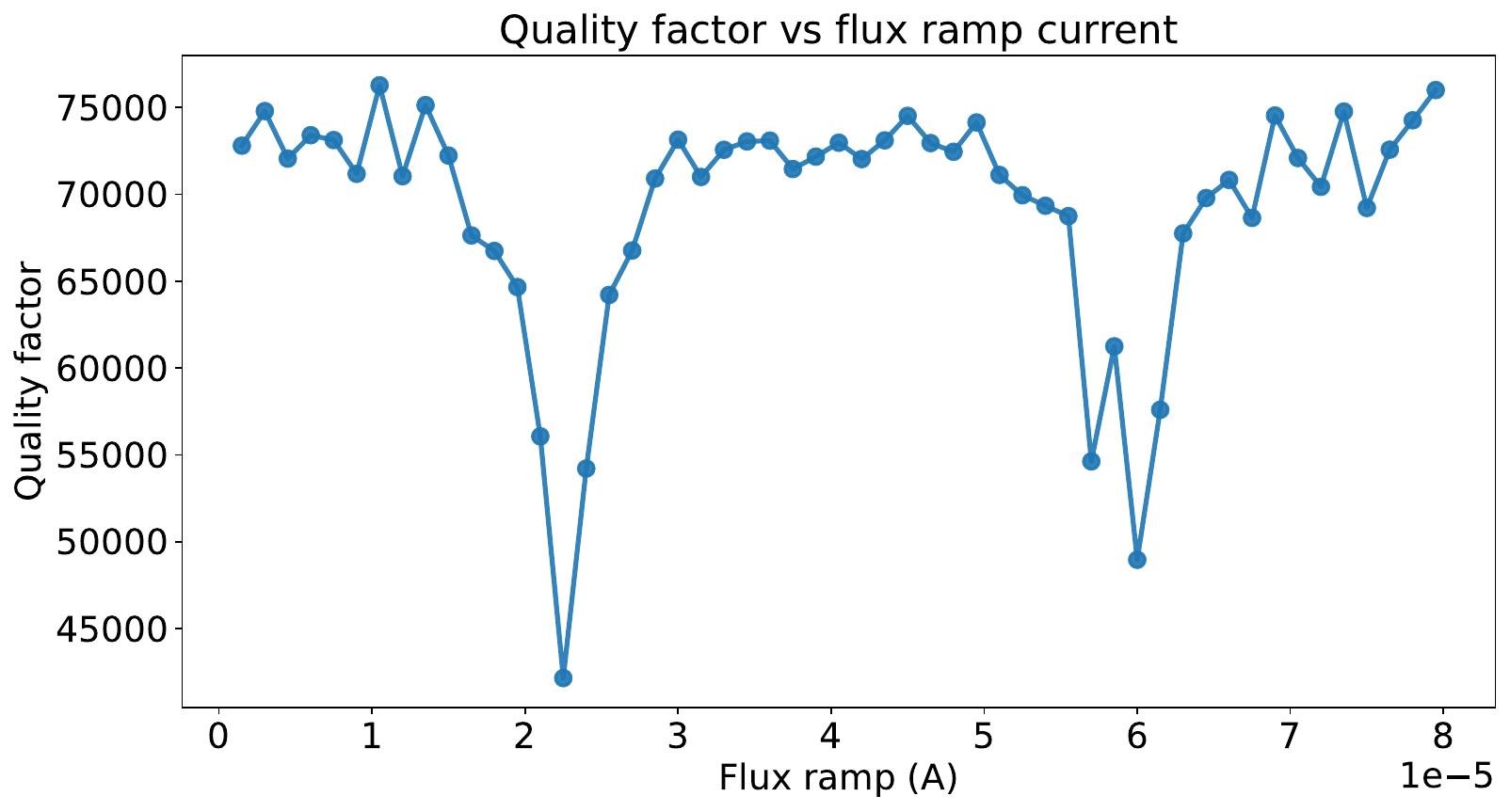}
    \caption{The dependence of the measured quality factor Q of ch01 on the flux ramp current. This characterization was conducted at 45 mK with an excitation power of -35 dBm.}
    \label{fig11}
\end{figure}

The internal quality factor of each channel has a periodic dependence on the flux ramp current. While the coupling quality factor is fixed, it is related to the coupling length and coupling spacing between the resonator and the feedline, and is not affected by the flux ramp current. Therefore, the quality factor of each channel also has a certain dependence on the flux ramp current. Fig. \ref{fig11} shows the dependence of the quality factor Q measured in ch01 on the flux ramp current. It can be seen that the dependence of Q on the flux ramp current is similar to that of Q$_i$ on the flux ramp current. Table \ref{table3}  presents a summary of the quality factors for the measured seven channels.

\begin{table}[h]
\centering
\caption{Summary of measured quality factors for the measured seven channels.}
\label{table3}
\begin{minipage}{0.7\textwidth}
\centering
\setlength{\tabcolsep}{0pt} 
\begin{tabular*}{\linewidth}{@{\extracolsep{\fill}}llll}
\toprule
Channel & Q$_i$ & Q$_c$ & Q \\
\midrule
1 & 137261 & 161098  & 73000 \\
2 & 123651 & 129359  & 64921 \\
3 & 46483 & 95269  & 30091 \\
4 & 100269 & 88366  & 40838 \\
5 & 33293 & 88358  & 22569 \\
6 & 118264 & 94487  & 45457 \\
7 & 33267 & 131713  & 26559 \\
\botrule
\end{tabular*}
\end{minipage}
\end{table}

\subsection{Flux modulation and readout noise measurement}\label{subsec3}

\begin{figure}[H]%
    \centering
    \includegraphics[width=0.8\textwidth]{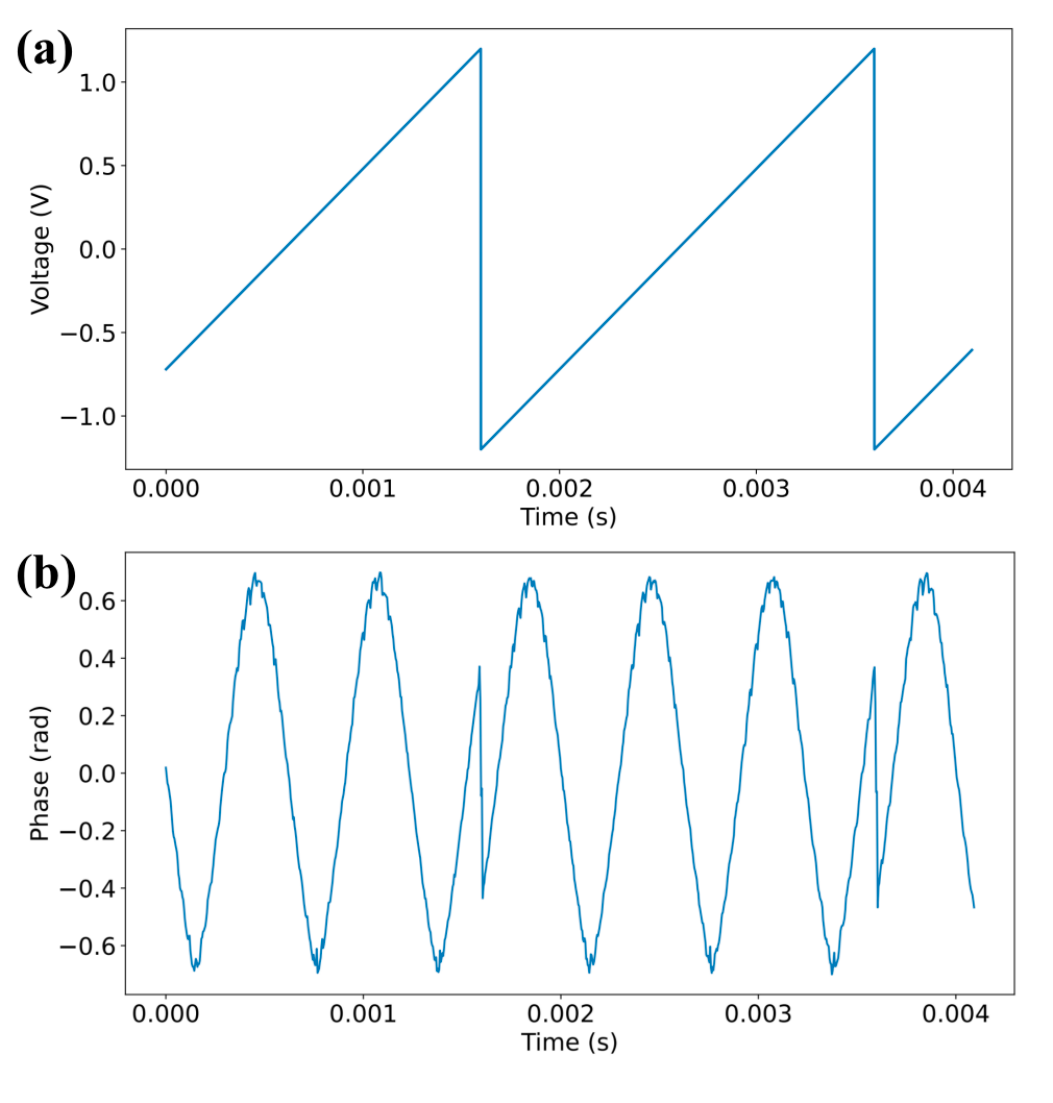}
    \caption{(a)Sawtooth wave signal; (b)Modulation curve}
    \label{fig12}
\end{figure}

\begin{figure}[h]%
    \centering
    \includegraphics[width=0.88\textwidth]{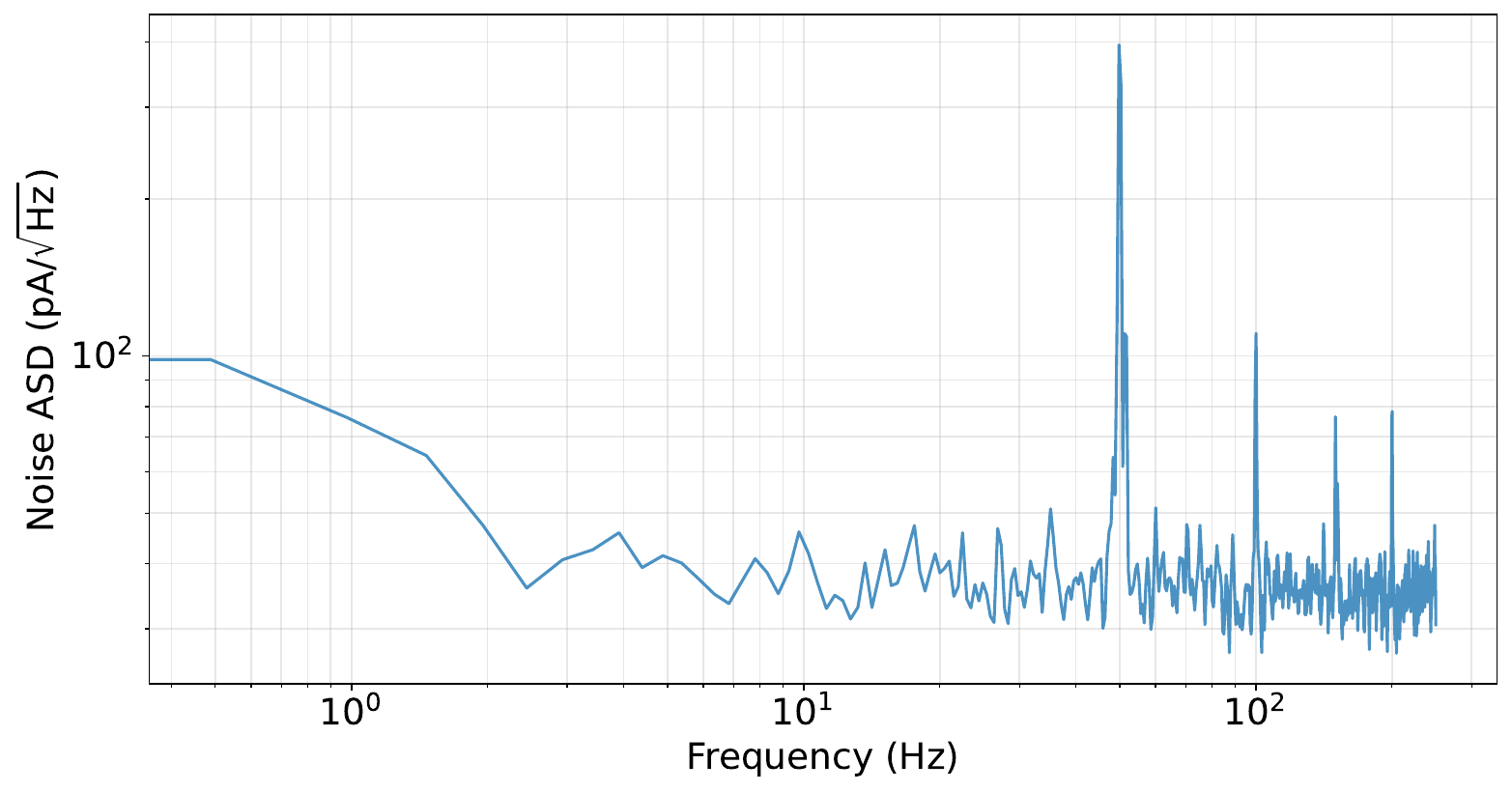}
    \caption{Measured equivalent noise current (NEI) of ch01}
    \label{fig13}
\end{figure}

By supplying a sawtooth wave signal(Figure \ref{fig12}(a)) to the flux ramp coil through a signal generator, the noise performance of the $\mu$Mux can be assessed. The excitation tone signals for reading out the seven channels of the $\mu$Mux prototype are generated and combined by the room-temperature electronics. The frequencies of the generated tone signals are derived from the fitted results of the measured S$_{21}$. The demodulated phase signal of the flux ramp modulation is illustrated in Figure \ref{fig12}(b). From this demodulated phase signal, we can extract the signal from the input coil of the $\mu$Mux. When no signal is input from the input coil, the demodulated signal can be utilized to derive the equivalent noise current using Fast Fourier Transform (FFT) analysis. The amplitude of the sawtooth wave signal used in our measurement is 2.4 V, corresponding to three magnetic flux quanta, with a frequency of 500 Hz. The excitation tone power is -17 dBm. The measured equivalent noise current power for ch01 is presented in Figure \ref{fig13}. It can be observed that the measured equivalent noise current (NEI) is 42 pA/$\sqrt{\text{Hz}}$ in the frequency range of 1--100 Hz, which is comparable to the NEI in the literature \citep{Dober-2017}.

\section{Conclusion}\label{sec6}

We fabricated a 32-channel microwave SQUID  multiplexer prototype and tested eight channels. Among these, seven channels exhibited magnetic flux-modulated responses. The measured equivalent noise current reached 42 pA/$\sqrt{Hz}$ and the quality factor was 73,000. While challenges remain, such as the lack of resonance points and inconsistencies in the critical current of the Josephson junctions, we plan to investigate and resolve these issues. By optimizing the etching conditions and the stress during film deposition, we anticipate that these problems will be addressed. Furthermore, we aim to fabricate and characterize a $\mu$Mux prototype with a higher multiplexing ratio, which will be utilized for the upgrade of AliCPT.

\backmatter

\bmhead*{Funding}

This work is supported by the National Key Research and Development Program of China (Grant No.2021YFC2203401), Scientific Instrument Developing Project of the Chinese Academy of Sciences (Grant No.YJKYYQ20190065), Youth Innovation Promotion Association CAS 2021011, Open Research Fund of the State Key Laboratory of Particle Astrophysics (Grant No. 2025000276).

\bibliography{sn-bibliography}

@article{Hubmayr-2022,
   author = {Hubmayr, J. and Ade, P. A. R. and Adler, A. and Allys, E. and Alonso, D. and Arnold, K. and et al.},
   title = {{Optical Characterization of OMT-Coupled TES Bolometers for LiteBIRD}},
   journal = {Journal of Low Temperature Physics},
   volume = {209},
   pages = {396–408},
   DOI = {10.1007/s10909-022-02808-7},
   year = {2022}
}

@article{Gualtieri-2016,
   author = {Gualtieri, R. and Battistelli, E.S. and Cruciani, A. and de Bernardis, P. and Biasotti, M. and Corsini, D. and Gatti, F. and Lamagna, L. and Masi, S},
   title = {{Multi-mode TES Bolometer Optimization for the LSPE-SWIPE Instrument}},
   journal = {Journal of Low Temperature Physics},
   volume = {184},
   pages = {527–533},
   DOI = {10.1007/s10909-015-1436-1},
   year = {2016}
}

@article{Barret-2023,
   author = {Barret, D. and Albouys, V. and Herder, JW.d. and Piro, L. and Cappi, M. and Huovelin, J. and Kelley, R. and Mas-Hesse, J.M. and Paltani, S. and Rauw, G. and Rozanska, A. and Svoboda,  J. and Wilms, J. and Yamasaki, N. and Audard, M. and Bandler, S. and et al.},
   title = {{The Athena X-ray Integral Field Unit: a consolidated design for the system requirement review of the preliminary definition phase}},
   journal = {Experimental Astronomy},
   volume = {55},
   pages = {373-426},
   DOI = {10.1007/s10686-022-09880-7},
   year = {2023}
}

@book{Clarke-2004,
  author		= "Clarke, J. and Braginski, A. I. ",
  title			= "The {SQUID} handbook: volume 1",
  publisher		= "",
  year			= "2004",
}

@book{Abazajian-2016,
    author = "Abazajian, Kevork N. and Adshead, Peter and Ahmed, Zeeshan and Allen, Steven W. and Alonso, David and Arnoldand, Kam S. and others",
    collaboration = "CMB-S4",
    title = "{CMB-S4 Science Book, First Edition}",
    publisher= "",
    doi = "10.2172/1352047",
    year = "2016"
}

@article{Doriese-2016,
   author = {Doriese, W. B. and Morgan, K. M. and Bennett, D. A. and others},
   title = {Developments in {Time-Division Multiplexing of X-ray Transition-Edge Sensors}},
   journal = {Journal of Low Temperature Physics},
   volume = {184},
   pages = {389-395},
   DOI = {10.1007/s10909-015-1373-z},
   year = {2016}
}

@article{Durkin-2021,
   author = {Durkin, Malcolm and Adams, Joseph S. and Bandler, Simon R. and Chervenak, James A. and Denison, Edward V. and Doriese, William B. and Duff, Shannon M. and Finkbeiner, Fred M. and Fowler, Joseph W. and Gard, Johnathon D. and Hilton, Gene C. and Hummatov, Ruslan and Irwin, Kent D. and Joe, Young Il and Kelley, Richard L. and Kilbourne, Caroline A. and Miniussi, Antoine R. and Morgan, Kelsey M. and O'Neil, Galen C. and Pappas, Christine G. and Porter, Frederick S. and Reintsema, Carl D. and Rudman, David A. and Sakai, Kazuhiro and Smith, Stephen J. and Stevens, Robert W. and Swetz, Daniel S. and Szypryt, Paul and Ullom, Joel N. and Vale, Leila R. and Wakeham, Nicholas},
   title = {Mitigation of {Finite Bandwidth Effects} in {Time-Division-Multiplexed} {SQUID} {Readout} of {TES} Arrays},
   journal = {IEEE Transactions on Applied Superconductivity},
   volume = {31},
   number = {5},
   pages = {1-5},
   DOI = {10.1109/TASC.2021.3065279},
   year = {2021}
}

@article{Goldfinger-2024,
   author = {Goldfinger, D. C. and Ahmed, Z. and Barron, D. R. and Doriese, W. B. and Durkin, M. and Filippini, J. P. and Haller, G. and Henderson, S. W. and Herbst, R. and Hubmayr, J. and Irwin, K. and Reese, B. and Sapozhnikov, L. and Thompson, K. L. and Ullom, J. and  Vissers, M. R.},
   title = {{End-to-End} {Modeling} of the {TDM} {Readout System} for {CMB-S4}},
   journal = {Journal of Low Temperature Physics},
   volume = {215},
   number = {3},
   pages = {143-152},
   DOI = {10.1007/s10909-024-03077-2},
   year = {2024}
}

@inproceedings{Henderson-2016,
   author = {Shawn W. Henderson and Jason R. Stevens and Mandana Amiri and Jason Austermann and James A. Beall and Saptarshi Chaudhuri and Hsiao-Mei Cho and Steve K. Choi and Nicholas F. Cothard and Kevin T. Crowley and Shannon M. Duff and Colin P. Fitzgerald and Patricio A. Gallardo and Mark Halpern and Matthew Hasselfield and Gene Hilton and Shuay-Pwu Patty Ho and Johannes Hubmayr and Kent D. Irwin and Brian J. Koopman and Dale Li and Yaqiong Li and Jeff McMahon and Federico Nati and Michael Niemack and Carl D. Reintsema and Maria Salatino and Alessandro Schillaci and Benjamin L. Schmitt and Sara M. Simon and Suzanne T. Staggs and Eve M. Vavagiakis and Jonathan T. Ward},
   title = {{Readout of two-kilopixel transition-edge sensor arrays for Advanced ACTPol}},
   volume = {9914},
   booktitle = {Millimeter, Submillimeter, and Far-Infrared Detectors and Instrumentation for Astronomy VIII},
   editor = {Wayne S. Holland and Jonas Zmuidzinas},
   organization = {International Society for Optics and Photonics},
   publisher = {SPIE},
   pages = {99141G},
   year = {2016},
   doi = {10.1117/12.2233895}
}

@article{Hattori-2016,
   author = {Hattori, K. and Akiba, Y. and Arnold, K. and Barron, D. and Bender, A. N. and Cukierman, A. and Haan, T. de and Dobbs, M. and Elleflot, T. and Hasegawa, M. and Hazumi, M. and Holzapfel, W. and Hori, Y. and Keating, B. and Kusaka, A. and Lee, A. and Montgomery, J. and Rotermund, K. and Shirley, I. and Suzuki, A. and Whitehorn, N.},
   title = {Development of {Readout Electronics for POLARBEAR-2 Cosmic Microwave Background Experiment}},
   journal = {Journal of Low Temperature Physics},
   volume = {184},
   pages = {512-518},
   DOI = {10.1007/s10909-015-1448-x},
   year = {2016}
}

@article{Vaccaro-2024,
   author = {Vaccaro, D. and Akamatsu, H. and Gottardi, L. and Wit, M. de and Bruijn, M. P. and Kuur, J. van der and Nagayoshi, K. and Taralli, E. and Ravensberg, K. and Gao, J.-R. and Herder, J. W. A. den},
   title = {Developments on frequency domain multiplexing readout for large arrays of transition-edge sensor X-ray micro-calorimeters},
   journal = {Journal of Low Temperature Physics},
   volume = {216},
   pages = {21-28},
   DOI = {10.1007/s10909-024-03099-w},
   year = {2024}
}

@article{Hartog-2018,
   author = {Hartog, R. den and Leeuwen, B.-J. van and Peille, P. and Kuur, J. van der and Ravera, L. and Loon, D. van and Jackson, B. and Herder, J.-W. den},
   title = {Performance of a state-of-the-art {DAC} system for {FDM} readout},
   journal = {Proc. SPIE 10699, Space Telescopes and Instrumentation 2018: Ultraviolet to Gamma Ray, 106994Q},
   DOI = {10.1117/12.2312793},
   year = {2018}
}

@article{Irwin-2010,
   author = {Irwin, K. D. and Niemack, M. D. and Beyer, J. and Cho, H. M. and Doriese, W. B. and Hilton, G. C. and others},
   title = {Code-division multiplexing of superconducting transition-edge sensor arrays},
   journal = {Superconductor Science and Technology},
   volume = {23},
   number = {3},
   pages = {034004},
   DOI = {10.1088/0953-2048/23/3/034004},
   year = {2010}
}

@article{Morgan-2016,
   author = {Morgan, K. M. and Alpert, B. K. and Bennett, D. A. and Denison, E. V. and Doriese, W. B. and Fowler, J. W. and Gard, J. D. and Hilton, G. C. and Irwin, K. D. and Joe, Y. I. and others},
   title = {Code-division multiplexed read-out of large arrays of tes microcalorimeters},
   journal = {Applied Physics Letters},
   volume = {109},
   pages = {112604},
   DOI = {10.1063/1.4962636},
   year = {2016}
}

@article{Irwin-2004,
   author = {Irwin, K. D. and Lehnert, K. W.},
   title = {Microwave {SQUID} multiplexer},
   journal = {Applied Physics Letters},
   volume = {85},
   number = {11},
   pages = {2107-2109},
   DOI = {10.1063/1.1791733},
   year = {2004}
}

@article{Cyndia-2023,
   author = {Yu, Cyndia and Ahmed, Z. and Frisch, J. C. and Henderson, S. W. and Silva-Feaver, M. and Arnold, K. and Brown, D. and Connors, J. and Cukierman, A. J. and D’Ewart, J. M. and Dober, B. J. and Dusatko, J. E. and Haller, G. and Herbst, R. and Hilton, G. C. and Hubmayr, J. and Irwin, K. D. and others},
   title = {{SLAC} microresonator {RF}({SMuRF}) electronics: A tone-tracking readout system for superconducting microwave resonator arrays},
   journal = {Review of Scientific Instruments},
   volume = {94},
   pages = {014712},
   DOI = {10.1063/5.0125084},
   year = {2023}
}

@article{Dober-2021,
   author = {Dober, B. and Ahmed, Z. and Arnold, K. and Becker, D. T. and Bennett, D. A. and Connors, J. A. and Cukierman, A. and D'Ewart, J. M. and Duff, S. M. and Dusatko, J. E. and Frisch, J. C. and Gard, J. D. and Henderson, S. W. and Herbst, R. and Hilton, G. C. and others},
   title = {A microwave {SQUID} multiplexer optimized for bolometric applications},
   journal = {Applied Physics Letters},
   volume = {118},
   pages = {062601},
   DOI = {10.1063/5.0033416},
   year = {2021}
}

@article{Groh-2025,
   author = {Groh, J. C. and Ahmed, Z. and Austermann, J. and Beall, J. and Daniel, D. and Duff, S. M. and Henderson, S. W. and Hubmayr,  J. and Lew, R. and Link, M. and Lucas, T. J. and Mates, J. A. B. and Silva-Feaver, M. and Singh, R. and Ullom, J. and Vale, L. and Van Lanen, J. and Vissers, M. and Yu, C.},
   title = {Demonstration of a 1820 channel multiplexer for transition-
edge sensor bolometers},
   journal = {Applied Physics Letters},
   volume = {127},
   pages = {152602},
   DOI = {10.1063/5.0290914},
   year = {2025}
}

@article{Cukierman-2019,
   author = {Cukierman, A. and Ahmed, Z. and Henderson, S. and Young, E. and Yu, C. and Barkats, D. and Brown, D. and Chaudhuri, S. and Cornelison, J. and D’Ewart, J. M. and Dierickx, M. and Dober, B. J. and Dusatko, J. and Fatigoni, S. and Filippini, J. P. and Frisch, J. C. and Haller, G. and Halpern, M. and Hilton, G. C. and Hubmayr, J. and Irwin, K. D. and others},
   title = {Microwave Multiplexing on the {Keck Array}},
   journal = {Journal of Low Temperature Physics},
   volume = {199},
   pages = {858-866},
   DOI = {10.1007/s10909-019-02296-2},
   year = {2019}
}

@article{Li-2020,
   author = {Li, Y. and Arnold, K. and Atkins, Z. and Bruno, S. M. and Cothard, N. F. and Dober, B. and Duell, C. J. and Duff, S. M. and Gallardo, P. A. and Healy, E. and Ho, S. P. and Hubmayr, J. and Keating, B. and Lee, A. T. and Mangu, A. and McCarrick, H. and Niemack, M. D. and Newburgh, L. and Raum, C. and Salatino, M. and Sasse, T. and Silva-Feaver, M. and Simon, S. M. and Staggs, S. and others},
   title = {{Assembly and Integration Process of the High-Density Detector Array Readout Modules for the Simons Observatory}},
   journal = {Journal of Low Temperature Physics},
   volume = {199},
   pages = {985-993},
   DOI = {10.1007/s10909-020-02386-6},
   year = {2020},
}

@article{Bennett-2019,
   author = {Bennett, D. and Mates, J. B. and Bandler, S. and Becker, D. and Fowler, J. and Gard, J. and Hilton, G. and Irwin, K. and Morgan, K. and Reintsema, C. and Sakai, Kazuhiro and Schmidt, D. and Smith, S. and Swetz, D. and Ullom, J. and Vale, L. and Wessels, A.},
   title = {{Microwave SQUID multiplexing for the Lynx x-ray microcalorimeter}},
   journal = {Journal of Astronomical Telescopes, Instruments, and Systems},
   volume = {5(2)},
   pages = {021007},
   DOI = {10.1117/1.JATIS.5.2.021007},
   year = {2019}
}

@article{Salatino-2020,
   author = {Salatino, M. and Austermann, J. and Thompson, K. L. and Ade, P. A. R. and Bai, X. and Beall, J. A. and Becker, D. T. and Cai, Y. and Chang, Z. and Chen, D. and Connors, J. and Chen, P. and Dober, B. and Delabrouille, J. and Duff, S. M. and Gao, G. and Givhan, R. C. and Ghosh, S. and Hilton, G. and Hu, B. and Hubmayr, J. and Karpel, E. and Kuo, C.-L. and Li, H. and Li, M. and Li, S.-Y. and Li, X. and Link, M. and Li, Y. and Liu, H. and Liu, L. and Liu, Y. and Lu, F. and Lucas, T. and Lu, X. and Mates, J. A. B. and others},
   title = {{The Design of the Ali CMB Polarization Telescope receiver}},
   journal={Millimeter, Submillimeter, and Far-Infrared Detectors and Instrumentation for Astronomy X},
   pages = {114532A},
   DOI = {10.1117/12.2560709},
   year = {2020}
}

@misc{inductex,
  title        = {InductEx Software},
  author       = {Fourie, Coenrad J.},
  organization = {Stellenbosch University},
  year         = {2015}, 
  howpublished = {\url{https://www.inductex.info}}
}

@phdthesis{Gao-2008-thesis,
   author = {Gao, J. S.},
   title = {{The Physics of Superconducting Microwave Resonators}},
   type = {Thesis},
   school = {California Institute of Technology},
   year = {2008}
}

@article{Kempf-2017,
   author = {Kempf, S. and Wegner, M. and Deeg, L. and Fleischmann, A. and Gastaldo, L. and Herrmann, F.  and Richter, D. and Enss, C.},
   title = {{Design, fabrication and characterization of a 64 pixel metallic magnetic calorimeter array with integrated, on-chip microwave SQUID multiplexer}},
   journal={Superconductor Science and Technology},
   volume = {30},
   pages = {065002},
   DOI = {10.1088/1361-6668/aa6d17},
   year = {2017}
}

@article{Dober-2017,
   author = {Dober, B. and Becker, D. T. and Bennett, D. A. and Bryan, S. A. and Duff, S. M. and Gard, J. D. and Hays-Wehle, J. P. and Hilton, G. C. and Hubmayr, J. and Mates,  J. A. B. and Reintsema, C. D. and Vale, L. R. and Ullom, J. N.},
   title = {{Microwave SQUID multiplexer demonstration for cosmic microwave background imagers}},
   journal={Applied Physics Letters},
   volume = {111},
   pages = {243510},
   DOI = {10.1063/1.5008527},
   year = {2017}
}

\end{document}